\documentclass[english,11pt,a4paper]{article}
\usepackage[T1]{fontenc}
\usepackage[latin9]{inputenc}
\usepackage{amsmath}
\usepackage{amsthm}
\usepackage{amssymb}
\usepackage{graphicx}

\makeatletter
\numberwithin{equation}{section}

\usepackage{jheppub}
\usepackage{slashed}
\usepackage[colorinlistoftodos, color=green!40, prependcaption]{todonotes}
\usepackage{orcidlink}
\usepackage{simplewick}

\makeatother

\usepackage{babel}
\begin{document}
\title{Impact of the cosmic neutrino background on long-range force searches} 

\abstract{
Light bosons can mediate long-range forces.
We show that light bosonic mediators interacting with a background medium, in particular,  with the cosmic neutrino background (C$\nu$B),  may induce medium-dependent masses
which could effectively screen long-range forces from detection. 
This leads to profound implications for long-range force searches in e.g.~the E\"ot-Wash,  MICROSCOPE, and lunar laser-ranging (LLR) experiments. 
For instance, we find that when the coupling of the mediator to neutrinos is above $3\times10^{-10}$ or $5\times10^{-13}$, bounds from LLR and experiments employing the Sun as an attractor, respectively, would be entirely eliminated. Larger values of the coupling can also substantially alleviate bounds from searches conducted at shorter distances.   
}

\author[a]{Garv Chauhan\,\orcidlink{0000-0002-8129-8034}}
     \emailAdd{gchauhan@vt.edu}
    \affiliation[a]{Center for Neutrino Physics, Department of Physics, Virginia Tech, Blacksburg, VA 24061, USA}
\author[b]{Xun-Jie Xu\,\orcidlink{0000-0003-3181-1386}}
     \emailAdd{xuxj@ihep.ac.cn}
    \affiliation[b]{Institute of High Energy Physics, Chinese Academy of Sciences, Beijing 100049, China}
\preprint{\today}  
\maketitle

\section{Introduction}

Among the four fundamental interactions in nature---gravity, electromagnetism,
and the strong and weak interactions---two of them are long-range
forces, observable at macroscopic scales. It is tempting to conceive
that new interactions arising from theories beyond the standard model
(SM) might generate extra long-range forces. If discovered, new long-range
forces would have paradigm-shifting implications for new physics explorations
and could even   lead to revolutionary applications in macroscopic-scale
technologies. 

Long-range forces are mediated by massless or sufficiently light mediators.
% with the mass below the inverse of the interaction range. 
Some examples
include light scalar or vector bosons in hidden sectors or gauge extensions~\cite{Heeck:2014zfa,Knapen:2017xzo,Xu:2020qek,Chauhan:2020mgv, Berbig:2020wve, Chauhan:2022iuh,Ghosh:2023ilw,Graf:2023dzf},  
axions~\cite{Hoedl:2011zz,Mantry:2014zsa,Okawa:2021fto}, majorons~\cite{Chikashige:1980ui},
dilatons~\cite{Taylor:1988nw,Ellis:1989as,Kaplan:2000hh}, dark matter~\cite{Costantino:2019ixl,Barbosa:2024tty}, or even
neutrinos~\cite{Feinberg:1968zz,Hsu:1992tg,Dzuba:2017cas,Orlofsky:2021mmy,Xu:2021daf,Coy:2022cpt,Ghosh:2022nzo}.
In recent years, the interest in light, feebly-interacting particles
has been persistently growing (see Refs.~\cite{Lanfranchi:2020crw,Antel:2023hkf}
for reviews), while long-range force searches (also known as {\it
fifth-force searches} in the literature) play a crucial role in constraining
ultra-light particles if their 
% de Broglie wavelengths 
Compton wavelengths 
can reach macroscopic
extent.  

Currently, the most sensitive searches are based on torsion-balance
tests of gravity~\cite{Wagner:2012ui}. New forces mediated by light
bosons with generic couplings to SM fermions might cause observable
deviations in the probe of  the weak equivalent principle (WEP) or
the inverse square law of gravity. At large distance scales comparable
to the Earth's radius (corresponding to a mediator mass below $10^{-14}$
eV), the E\"ot-Wash experiment has excluded the couplings to electrons
and baryons above $10^{-23}\sim10^{-24}$. If such a small coupling
arises from high-energy theories (e.g.~generated by heavy loops~\cite{Xu:2020qek,Chauhan:2020mgv}),
then the extremely small value can be reinterpreted as a probe of
new physics at very high energy scales. Other experimental probes
including lunar laser-ranging (LLR)~\cite{Williams:1995nq,Williams:2004qba,Merkowitz:2010kka,Murphy:2013qya,Viswanathan:2017vob},
 the detection of gravitational waves at LIGO/Virgo~\cite{TheLIGOScientific:2017qsa,Abbott:2020khf,Croon:2017zcu,Kopp:2018jom,Dror:2019uea}, 
and the very recent MICROSCOPE experiment~\cite{MICROSCOPE:2022doy}
have their respective advantages in long-range force searches. 

Given the remarkably high sensitivity of these experiments to new
long-range forces, it is important to ask: to what extent are these
bounds valid and what factors may significantly alter the bounds?

Taking the example of electromagnetism,  the Coulomb potentials of
electrons and nuclei always cancel each other at large scales in electrically
neutral matter, and electromagnetic waves can often be screened by
metals. Consequently, the photon and photon-like particles\footnote{By ``photon-like'', we mean that the effective couplings to electrons
and nuclei are proportional to their electric charges. The dark photon
and the $L_{\mu}-L_{\tau}$ gauge boson with loop-induced couplings
to electrons and baryons are in this category. } are unable to cause observable effects in these experiments. For
axions and majorons which are pseudoscalars, there are similar cancellations
between particles of opposite spin directions in unpolarized matter.
 For light scalar or vector bosons, in general the forces on individual
particles can be added coherently without cancellation. However, such
bosons could still evade the aforementioned constraints if their propagation
in a medium behaves differently from that in the vacuum. 

 In this study, we investigate potential medium effects that may
exert a significant influence on the searches for long-range forces.
We consider a light scalar mediator $\phi$ that is generically coupled
to all SM fermions including neutrinos, and show that the propagation
of $\phi$ in a medium such as  normal matter or the cosmic neutrino
background (C$\nu$B) could be modified by coherent forward scattering
with medium particles, generating a medium-dependent mass for $\phi$.
Such a mass could shorten the interaction range and alleviate some
experimental bounds on long-range forces. We examine this effect and
find that the medium effect caused by normal matter of the Earth or
the Sun is negligible in experiments employing them as the attractor.
On the other hand, the C$\nu$B which is ubiquitous and cannot be
evacuated in any experiments, can cause a considerably strong medium
effect due to the lightness of neutrinos. We show that within a broad
range of the neutrino coupling allowed by cosmology, existing bounds
on long-range forces can be altered by the C$\nu$B very significantly.

This work is organized as follows. In Sec.~\ref{sec:framework},
we briefly review the formalism for generating long-range forces in the 
vacuum and elucidate the medium effect to be taken into account in the formalism. 
The medium effect can be included by solving a modified field 
equation, which is solved in Sec.~\ref{sec:solution} assuming a spherically symmetric configuration.   Then we inspect to what extent
the experimental bounds on long-range forces can be altered by the
C$\nu$B in Sec.~\ref{sec:result} and draw our conclusions in Sec.~\ref{sec:Conclusions}.
Some minus sign issues are addressed in Appendix~\ref{sec:minus-signs}
and a derivation of the medium effect is presented in Appendix~\ref{sec:Rederivation}. 

\section{Long-range forces and the medium effect\label{sec:framework}}

\subsection{Long-range forces in the vacuum}

Let us briefly review the generation of long-range forces from light
bosons, mainly following the formalism in Refs.~\cite{Smirnov:2019cae,Babu:2019iml}.
We start by considering the following Lagrangian:

\begin{equation}
{\cal L}=\frac{1}{2}\left(\partial\phi\right)^{2}-\frac{1}{2}m_{\phi}^{2}\phi^{2}+\overline{\psi}i\slashed{\partial}\psi-m_{\psi}\overline{\psi}\psi-y_{\psi}\phi\overline{\psi}\psi\thinspace,\label{eq:-1}
\end{equation}
where $\phi$ is a light scalar and $\psi$ denotes a generic fermion,
with their respective masses $m_{\phi}$ and $m_{\psi}$. The Lagrangian
leads to the following equations of motion (EOMs):
\begin{align}
i\slashed{\partial}\psi-(m_{\psi}+y_{\psi}\phi)\psi & =0\thinspace,\label{eq:-2}\\
(\partial^{2}+m_{\phi}^{2})\phi+y_{\psi}\overline{\psi}\psi & =0\thinspace.\label{eq:-3}
\end{align}
Eq.~\eqref{eq:-2} implies that in the presence of a background $\phi$
field, the mass of $\psi$ is essentially shifted to $m_{\psi}+y_{\psi}\phi$.
For nonrelativistic $\psi$ particles, this is equivalent to shifting
its energy by $y_{\psi}\phi$. Thus one can introduce an effective
potential to account for the influence of $\phi$ on $\psi$:
\begin{equation}
V=y_{\psi}\phi\thinspace.\label{eq:-7}
\end{equation}
When $\phi$ has a nonzero gradient ($\nabla\phi\neq0$), it causes
a force, $F=y_{\psi}\nabla\phi$, acting on $\psi$ particles. For
relativistic $\psi$ particles, the mass shift is not fully equivalent
to the energy shift, causing a reduction of the Yukawa force. We refer
to Ref.~\cite{Smirnov:2022sfo} for detailed discussions on this
issue. 

For a bulk of $\psi$ particles statically distributed in space, we
replace $\overline{\psi}\psi$ in Eq.~\eqref{eq:-3} with its ensemble
average $\langle\overline{\psi}\psi\rangle\approx n_{\psi}$ where
$n_{\psi}$ is the number density of $\psi$ particles. Since it is
static, we can neglect the temporal derivative in Eq.~\eqref{eq:-3}
and write it as
\begin{equation}
\left[\nabla^{2}-m_{\phi}^{2}\right]\phi=y_{\psi}n_{\psi}\ \text{(non-relativistic)}.\label{eq:-5}
\end{equation}

Assuming the distribution is spherically symmetric and using spherical
coordinates, we have $\nabla^{2}\phi=\frac{\partial^{2}\phi}{\partial r^{2}}+2\frac{\partial\phi}{r\partial r}=\frac{u''(r)}{r}$
where $u(r)\equiv r\phi(r)$. This allows us to further rewrite Eq.~\eqref{eq:-5}
as
\begin{equation}
\left[\frac{d^{2}}{dr^{2}}-m_{\phi}^{2}\right]u(r)=ry_{\psi}n_{\psi}(r)\thinspace,\label{eq:-8}
\end{equation}
which can be solved analytically for some simple forms of $n_{\psi}(r)$,
including e.g.~constant or piecewisely constant $n_{\psi}$, or the
exponential form $n_{\psi}\propto e^{-r\kappa}$~\cite{Smirnov:2019cae}.
In particular, if $n_{\psi}=N_{\psi}\delta^{3}(\mathbf{r})$, the
solution is $u(r)=-y_{\psi}N_{\psi}e^{-m_{\phi}r}/(4\pi)$, resulting
in the well-known Yukawa potential:
\begin{equation}
V=-\frac{y_{\psi}^{2}N_{\psi}}{4\pi r}e^{-m_{\phi}r}\thinspace.\label{eq:-9}
\end{equation}
The force arising from Eq.~\eqref{eq:-9}, $F=\nabla V$,  is considered
long-range if $m_{\phi}\ll r^{-1}$, in which case the exponential
suppression can be neglected. As is well known, the potential is negative,
implying that the Yukawa force is attractive. For vector interactions,
one can derive a similar potential which is positive, corresponding
to a repulsive force. 

\subsection{The medium effect\label{subsec:The-medium-effect}}

In a medium, the propagation of the $\phi$ field could be significantly
altered, similar to various known medium effects on electromagnetic
waves. For instance, in plasma, due to interactions with free charged
particles, the photon acquires an effective mass known as the plasmon
mass. In cold (non-ionized) gas such as the air,  where electrons
and nuclei are bound into atoms or molecules, the response of atomic
or molecular electric dipoles to electromagnetic waves slightly slows
down the propagation of light by a factor of $1/n$ with $n$ the
refraction index of the medium. 

For $\phi$ propagating through a medium containing a large number
of free $\psi$ particles, the medium effect is similar to the plasmon
mass, giving rise to the following mass correction for the $\phi$
field:
\begin{equation}
m_{\phi}^{2}\to m_{\phi}^{2}+y_{\psi}^{2}\frac{\tilde{n}_{\psi}}{m_{\psi}}\thinspace,\label{eq:-6}
\end{equation}
where
\begin{equation}
\tilde{n}_{\psi}\equiv\int\frac{d^{3}\mathbf{p}}{(2\pi)^{3}}\frac{m_{\psi}}{E_{\mathbf{p}}}f(\mathbf{p})\thinspace,\label{eq:-10}
\end{equation}
with $f(\mathbf{p})$ the phase space distribution function of $\psi$
particles. We refer to $\tilde{n}_{\psi}$ in Eq.~\eqref{eq:-10}
as {\it the effective number density}, in contrast to the standard
number density: 
\begin{equation}
n_{\psi}\equiv\int\frac{d^{3}\mathbf{p}}{(2\pi)^{3}}f(\mathbf{p})\thinspace.\label{eq:-11}
\end{equation}

Eq.~\eqref{eq:-6} can be derived from finite-temperature or finite-density
field theory~\cite{Babu:2019iml}. However, such a derivation involves
the underlying assumption that the medium is in thermal and chemical
equilibrium, which implies that $f(\mathbf{p})$ should take the Fermi-Dirac
distribution. In fact, Eq.~\eqref{eq:-6} also holds for general forms
of $f(\mathbf{p})$,  not necessarily limited to thermal distributions,
since it is a consequence of coherent forward scattering of $\phi$
with the medium particles. This is again similar to the photon, for
which the plasmon mass and the refraction index are known to arise from
coherent forward scattering. In Appendix~\ref{sec:Rederivation},
we rederive Eq.~\eqref{eq:-6} from the theory of coherent forward
scattering, following similar calculations in Refs.~\cite{Li:2023vpv,Wu:2024fsf}
for photon-dark photon or photon-axion conversions. This justifies
the use of Eq.~\eqref{eq:-6} for non-thermal distributions such as
the C$\nu$B. 

Let us note here that the mass correction in Eq.~\eqref{eq:-6} is
a local rather than global effect. In a density-varying medium, one
should only apply the mass correction at the  EOM level, not the
Yukawa potential in Eq.~\eqref{eq:-9} which is obtained by assuming
constant $m_{\phi}$. The actual potential should be determined by
solving the modified EOM:
\begin{equation}
\left[\nabla^{2}-m_{\phi}^{2}-y_{\psi}^{2}\frac{\tilde{n}_{\psi}}{m_{\psi}}\right]\phi=y_{\psi}\tilde{n}_{\psi}\thinspace.\label{eq:-5-1}
\end{equation}
Note that comparing to Eq.~\eqref{eq:-5}, the right-hand side has
also been changed. The right-hand side is changed from $y_{\psi}n_{\psi}$
to $y_{\psi}\tilde{n}_{\psi}$ because $\langle\overline{\psi}\psi\rangle\approx n_{\psi}$
is only applicable to nonrelativistic distributions while $\langle\overline{\psi}\psi\rangle=\tilde{n}_{\psi}$
holds for general cases including relativistic distributions~\cite{Smirnov:2022sfo}. 

By definition, $n_{\psi}$ and $\tilde{n}_{\psi}$ in Eqs.~\eqref{eq:-10}
and \eqref{eq:-11} do not include antiparticles. For antiparticles,
we define similar quantities, $n_{\overline{\psi}}$ and $\tilde{n}_{\overline{\psi}}$.
 When including antiparticles, there are a few subtle minus signs
to be discussed in  Appendix~\ref{sec:minus-signs} and the conclusion
is that antiparticles contribute positively to Eq.~\eqref{eq:-6},
i.e.~there is no cancellation between particles and antiparticles
in the medium effect. Therefore, the contribution of antiparticles
can be included by
\begin{equation}
\tilde{n}_{\psi}\to\tilde{n}_{\psi}+\tilde{n}_{\overline{\psi}}\thinspace.\label{eq:-25}
\end{equation}
In addition, Eq.~\eqref{eq:-5-1} can be straightforwardly generalized
to include multiple fermionic species that are coupled to $\phi$.
Altogether, the most general equation that determines the $\phi$
field in the presence of multiple species of medium particles together
with antiparticles reads:
\begin{equation}
\left[\nabla^{2}-\tilde{m}_{\phi}^{2}\right]\phi=\sum_{\psi}y_{\psi}\left(\tilde{n}_{\psi}+\tilde{n}_{\overline{\psi}}\right),\label{eq:-5-2}
\end{equation}
where
\begin{equation}
\tilde{m}_{\phi}^{2}\equiv m_{\phi}^{2}+\sum_{\psi}y_{\psi}^{2}\frac{\tilde{n}_{\psi}+\tilde{n}_{\overline{\psi}}}{m_{\psi}}\thinspace.\label{eq:-5-3}
\end{equation}

%{\color{red}{[}add comments on why $m_{\psi}\to m_{\psi}+y_{\psi}\phi$
%is not included in Eq.~\eqref{eq:-5-1}. Conclusion: effect too small.{]}}

\subsection{On the effective number density $\tilde{n}_{\psi}$}

In the nonrelativistic limit, the effective number density $\tilde{n}_{\psi}$
defined in Eq.~\eqref{eq:-10} simply reduces to $n_{\psi}$ because
$m_{\psi}/E_{\mathbf{p}}\approx1$. For relativistic distributions,
$\tilde{n}_{\psi}$ is smaller than $n_{\psi}$ and their ratio $\tilde{n}_{\psi}/n_{\psi}$
is roughly suppressed by $m_{\psi}/\langle E_{\mathbf{p}}\rangle$
where $\langle E_{\mathbf{p}}\rangle$ denotes the mean energy of
$\psi$ particles. 

In general, by computing the integral in Eq.~\eqref{eq:-10} for a
given $f(\mathbf{p})$, one obtains
\begin{equation}
\tilde{n}_{\psi}\approx\begin{cases}
n_{\psi} & (\text{nonrelativistic})\\
\xi m_{\psi}n_{\psi}^{2/3} & (\text{relativistic})
\end{cases}\thinspace,\label{eq:-13}
\end{equation}
where $\xi$ is an ${\cal O}(1)$ coefficient depending on the
shape of $f(\mathbf{p})$. For the Fermi-Dirac (FD) distribution with
zero chemical potential, 
\begin{equation}
\xi=\frac{1}{6}\left(\frac{\pi^{2}}{6\zeta(3)}\right)^{2/3}g_{i}^{1/3}\approx0.2054g_{i}^{1/3}\thinspace,\label{eq:-40}
\end{equation}
%$\xi\approx0.2054g_{i}^{1/3}$ 
with $g_{i}$
the number of  internal degrees of freedom: $g_{i}=2$ for electrons
and  $g_{i}=1$ for Majorana neutrinos. 
%{\color{red}{[}add $\zeta(3)....${]}}
For
the FD distribution distribution with zero temperature but a finite
chemical potential (i.e. the degenerate limit), $\xi\approx0.3848g_{i}^{1/3}$.
The Maxwell-Boltzmann distribution would lead to $\xi\approx0.2331g_{i}^{1/3}$. 

Note that the standard cosmic neutrino background (C$\nu$B) with
nonzero neutrino masses is not exactly in the FD distribution (see
e.g.~\cite{Esteban:2021ozz} for detailed discussions): 
\begin{equation}
f_{\text{C}\nu\text{B}}(p)=\frac{1}{e^{p/T}+1}\neq\frac{1}{e^{E_{p}/T}+1}\thinspace.\label{eq:-12}
\end{equation}
But for neutrino masses well below $1.9\ \text{K}\approx1.6\times10^{-4}\ \text{eV}$
(possible for the lightest neutrino mass eigenstate), the difference
is negligible and one can take $\xi\approx0.2054g_{i}^{1/3}$. 

The neutrino mass squared differences in neutrino oscillations have
been well measured~\cite{Zyla:2020zbs}:
\begin{equation}
|\Delta m_{31}^{2}|\approx 2.5\times10^{-3}\ \text{eV}^{2}\thinspace,\ \Delta m_{21}^{2}\approx7.4\times10^{-5}\ \text{eV}^{2}\thinspace,\label{eq:-14}
\end{equation}
which implies that two of the mass eigenstates are heavier than $\sqrt{|\Delta m_{31}^{2}|}\approx0.05$
eV or $\sqrt{\Delta m_{21}^{2}}\approx0.009$ eV. Therefore, among
three neutrino mass eigenstates, at least two of them should be nonrelativistic
today. 

%{\color{red}{[}Comments on Majorana vs Dirac?{]}}

\section{The spherically symmetric solution\label{sec:solution}}

The differential equation \eqref{eq:-5-1} can be solved analytically
if the effective number density is spherically symmetric with the
following form: 
\begin{equation}
\tilde{n}_{\psi}(r)=\begin{cases}
\tilde{n}_{\psi0} & \text{for }r\leq R\\
0 & \text{for }r>R
\end{cases}\thinspace,\label{eq:-28}
\end{equation}
where $\tilde{n}_{\psi0}$ is a constant and $R$ is the radius of
the spherical distribution.  By solving Eq.~\eqref{eq:-5-1} in the
$r\leq R$ and $r>R$ regimes and requiring that $\phi$ and $d\phi/dr$
are continuous at $r=R$, we obtain the solution
\begin{equation}
\phi=\begin{cases}
\frac{y_{\psi}\tilde{n}_{\psi0}}{\tilde{m}_{\phi}^{2}}\left[-1+\frac{C_{{\rm in}}}{r}\sinh(\tilde{m}_{\phi}r)\right] & \text{for }r\leq R\\[2mm]
\frac{y_{\psi}\tilde{n}_{\psi0}}{\tilde{m}_{\phi}^{2}}\cdot\frac{C_{{\rm out}}}{r}e^{-m_{\phi}(r-R)} & \text{for }r>R
\end{cases}\thinspace,\label{eq:-29}
\end{equation}
where $\tilde{m}_{\phi}$ denotes the effective mass within the sphere
($\tilde{m}_{\phi}^{2}=m_{\phi}^{2}+y_{\psi}^{2}\tilde{n}_{\psi0}/m_{\psi}$)
and 
\begin{align}
C_{{\rm in}} & =\frac{m_{\phi}R+1}{\tilde{m}_{\phi}\cosh(\tilde{m}_{\phi}R)+m_{\phi}\sinh(\tilde{m}_{\phi}R)}\thinspace,\label{eq:-27}\\
C_{{\rm out}} & =\frac{\sinh(\tilde{m}_{\phi}R)-\tilde{m}_{\phi}R\cosh(\tilde{m}_{\phi}R)}{m_{\phi}\sinh(\tilde{m}_{\phi}R)+\tilde{m}_{\phi}\cosh(\tilde{m}_{\phi}R)}\thinspace.\label{eq:-30}
\end{align}
Here we would like to discuss an interesting limit: $m_{\phi}\to0$,
i.e.~$\phi$ is massless in the vacuum but acquires a mass within the
sphere due to the medium. In this limit, Eq.~\eqref{eq:-29} reduces
to
\begin{equation}
\phi=\begin{cases}
-\frac{y_{\psi}\tilde{n}_{\psi0}}{\tilde{m}_{\phi}^{2}}\left[1-\frac{\sinh(\tilde{m}_{\phi}r)}{\tilde{m}_{\phi}r\cosh(\tilde{m}_{\phi}R)}\right] & \text{for }r\leq R\\[2mm]
-\frac{y_{\psi}\tilde{n}_{\psi0}}{\tilde{m}_{\phi}^{2}}\left[1-\frac{\tanh(\tilde{m}_{\phi}R)}{\tilde{m}_{\phi}R}\right]\frac{R}{r} & \text{for }r>R
\end{cases}\thinspace,\ \ (m_{\phi}\to0)\thinspace.
\end{equation}
Obviously, the $r>R$ part of the solution behaves like an infinitely
long-range force, as can be seen from its $1/r$ dependence.
If the medium effect is weak ($\tilde{m}_{\phi}R\ll1$), one can further
expand it in terms of $x\equiv\tilde{m}_{\phi}R$: 
\begin{equation}
\phi=\begin{cases}
y_{\psi}\tilde{n}_{\psi0}\left[\frac{r^{2}-3R^{2}}{6}+\frac{x^{2}\left(r^{2}-5R^{2}\right)^{2}}{120R^{2}}+{\cal O}(x^{4})\right] & \text{for }r\leq R\\[4mm]
-\frac{y_{\psi}\tilde{n}_{\psi0}R^{3}}{3r}\left[1-\frac{2x^{2}}{5}+{\cal O}(x^{4})\right] & \text{for }r>R
\end{cases}\thinspace,\ \ (m_{\phi}\to0)\thinspace.
\end{equation}
Here the first terms in squared brackets correspond to known results
of the Coulomb potential, while the second terms represent the leading-order
medium effect. \\

In the presence of multiple species of medium particles coupled to
$\phi$, one needs to solve the more general differential equation
\eqref{eq:-5-2}. Note that due to the mass correction term in Eq.~\eqref{eq:-5-3},
Eq.~\eqref{eq:-5-2} cannot be solved by linearly adding the solutions
in Eq.~\eqref{eq:-29} for each species. Unlike Eq.~\eqref{eq:-5}
for which the solution scales linearly with respect to  the number
density, Eqs.~\eqref{eq:-5-1} and \eqref{eq:-5-2} exhibit nonlinearity
when the medium effect is significant. 

Nevertheless, Eq.~\eqref{eq:-29} can still be practically useful
under certain approximations and assumptions. Consider for instance
the Earth as a spherically symmetric object. Assuming that its matter
density and chemical composition are homogeneous, the above solution can be used by replacing $y_{\psi}\tilde{n}_{\psi0}\to\sum_{\psi}y_{\psi}(\tilde{n}_{\psi0}+\tilde{n}_{\overline{\psi}0})$
in Eq.~\eqref{eq:-29} and taking 
$\tilde{m}_{\phi}^{2}=m_{\phi}^{2}+\sum_{\psi}y_{\psi}^{2}(\tilde{n}_{\psi0}+\tilde{n}_{\overline{\psi}0})/m_{\psi}$,
where the summation $\sum_{\psi}$ goes over all ingredients  of
the Earth. Furthermore, adding the C$\nu$B to this problem can also
be solved analytically, assuming that the C$\nu$B is homogeneous
and extends infinitely in space. Under this assumption, the medium
effect of C$\nu$B is a constant shift added to $m_{\phi}^{2}$. 
%{\color{red}{[}comment on ``not-necessarily-a-constant''{]}}

Another example is the Yukawa force between two celestial bodies.
Assuming that each celestial body is a spherically symmetric object
with a homogeneous matter density and the two bodies are well separated,
one can first solve their respective $\phi$ field equations:
\begin{align}
\left[\nabla^{2}-\tilde{m}_{\phi1}^{2}\right]\phi_{1} & =y_{\psi}\tilde{n}_{\psi1}\thinspace,\label{eq:-31}\\
\left[\nabla^{2}-\tilde{m}_{\phi2}^{2}\right]\phi_{2} & =y_{\psi}\tilde{n}_{\psi2}\thinspace,\label{eq:-32}
\end{align}
where $\phi_{i}$ ($i=1$, 2) denotes the field strength generated
by the $i$-th celestial body and $\tilde{m}_{\phi i}^{2}=m_{\phi}^{2}+y_{\psi}^{2}\tilde{n}_{\psi i}/m_{\psi}$.
Here for simplicity we assume that they all consist of a single species
of $\psi$ particles.  Summing the two solutions $\phi_{1}$ and
$\phi_{2}$ together cannot generate an exact solution of the combined
system, but $\phi=\phi_{1}+\phi_{2}$ satisfies
\begin{equation}
\left[\nabla^{2}-\tilde{m}_{\phi}^{2}\right]\phi=y_{\psi}\tilde{n}_{\psi1}+y_{\psi}\tilde{n}_{\psi2}+y_{\psi}^{2}\frac{\tilde{n}_{\psi2}\phi_{1}+\tilde{n}_{\psi1}\phi_{2}}{m_{\psi}}\thinspace,\label{eq:-33}
\end{equation}
where $\tilde{m}_{\phi}^{2}=m_{\phi}^{2}+y_{\psi}^{2}(\tilde{n}_{\psi1}+\tilde{n}_{\psi2})/m_{\psi}$
and the last term in Eq.~\eqref{eq:-33} comes from cross terms in
$\tilde{m}_{\phi}^{2}\phi$. Compared to $y_{\psi}\tilde{n}_{\psi1}$
or $y_{\psi}\tilde{n}_{\psi2}$, the last term is suppressed by a
factor of $y_{\psi}\phi_{i}/m_{\psi}$ which is $\ll1$ as long as
the mass shift of $\psi$ {[}see the discussion above Eq.~\eqref{eq:-7}{]}
is small. Hence $\phi=\phi_{1}+\phi_{2}$ can be viewed as an approximate
solution of the field equation for the two-body system.

\section{Altering the experimental bounds on long-range forces\label{sec:result}}

A variety of experiments measuring gravitational effects or testing
gravitational laws can also be utilized to search for extra long-range
forces, as has been conducted by a number of experimental groups~\cite{Hoskins:1985tn,Schlamminger:2007ht,Adelberger:2009zz,Wagner:2012ui,Yang:2012zzb,Tan:2016vwu,Tan:2020vpf,fifth-force,Lee:2020zjt}. In recent years, the detection of gravitational waves from
black hole or neutron star binary mergers also offers an important
avenue for probing long-range forces~\cite{Croon:2017zcu,Kopp:2018jom,Dror:2019uea}. % {[}xxx ???{]}. 

We focus our analyses on experimental bounds derived from measuring
gravitational effects of the most well-understood celestial bodies
nearby, namely the Sun, Earth, and Moon. With these celestial bodies
as attractors, new long-range forces can be severely constrained by
measuring possible acceleration differences between two test bodies
composed of different materials (e.g. Be, Ti, Al, Pt), known as the
test of the weak equivalence principle (WEP). Such a test has been
conducted by the Princeton~\cite{Roll:1964rd} and Moscow~\cite{1972JETP} groups utilizing
the Sun as the attractor and subsequently by the E\"ot-Wash group~\cite{Schlamminger:2007ht,Wagner:2012ui} using the Earth as the attractor. In addition to the WEP
test, there is another interesting type of experimental probe---the
lunar laser-ranging (LLR) experiments~\cite{Williams:1995nq,Williams:2004qba,Merkowitz:2010kka,Murphy:2013qya,Viswanathan:2017vob}
which are capable to measure anomalous precession of the lunar orbit
and hence sensitive to new forces that modify the inverse square law. Very recently, the MICROSCOPE experiment~\cite{MICROSCOPE:2022doy} deployed test masses orbiting the Earth in a drag-free satellite and achieved an unprecedented precision of measuring WEP violation. 
%{\color{red}{[}add comments on MICROSCOPE, Ref.~\cite{MICROSCOPE:2022doy}{]}}

\begin{figure}
\centering

\includegraphics[width=0.5\textwidth]{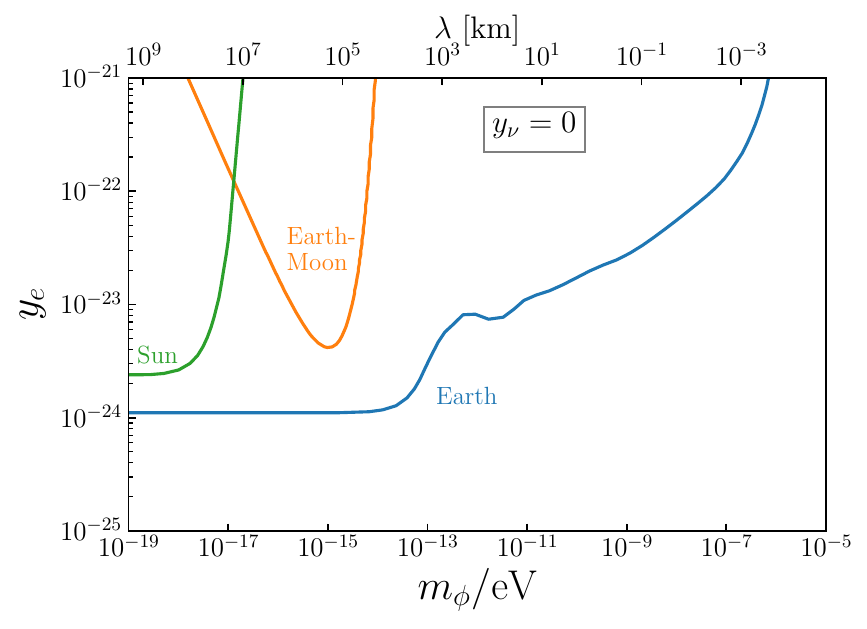}\includegraphics[width=0.5\textwidth]{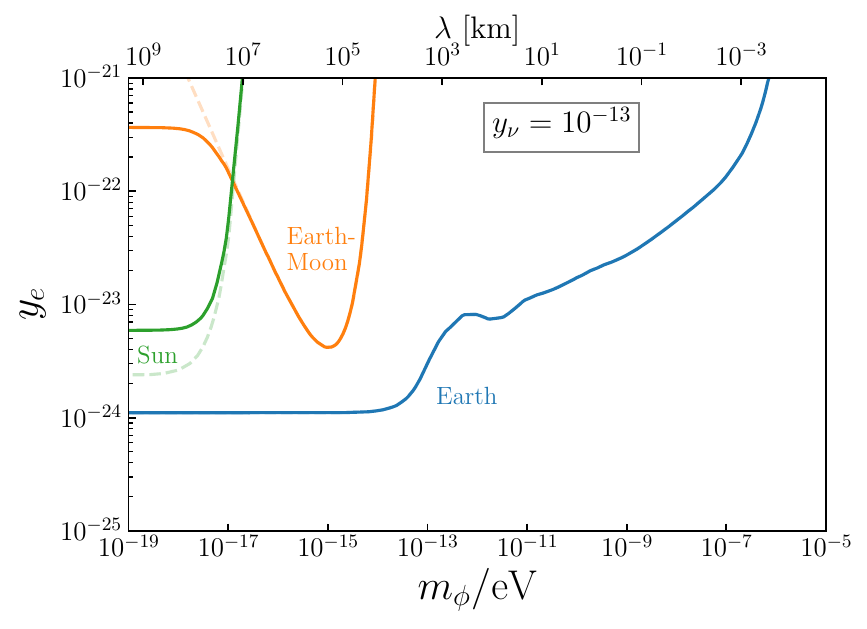}

\includegraphics[width=0.5\textwidth]{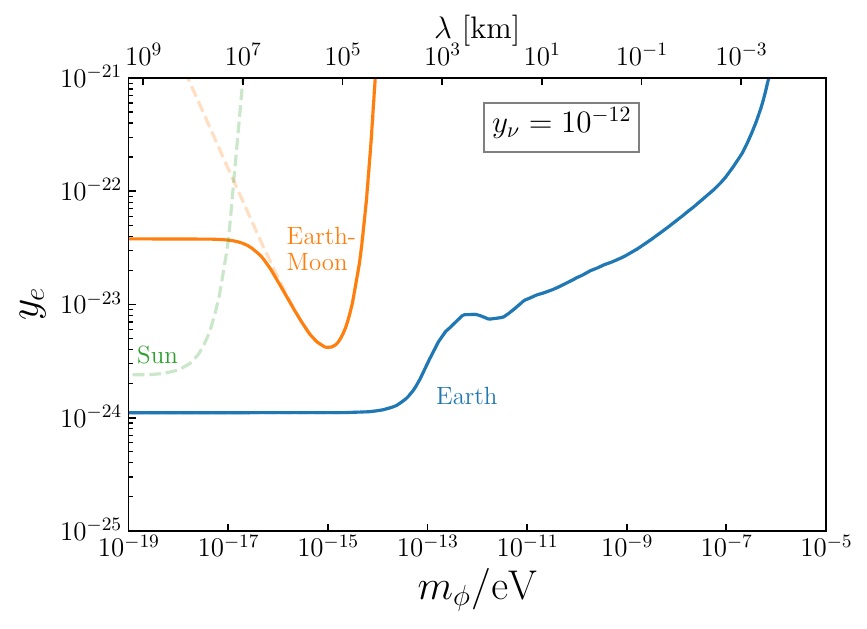}\includegraphics[width=0.5\textwidth]{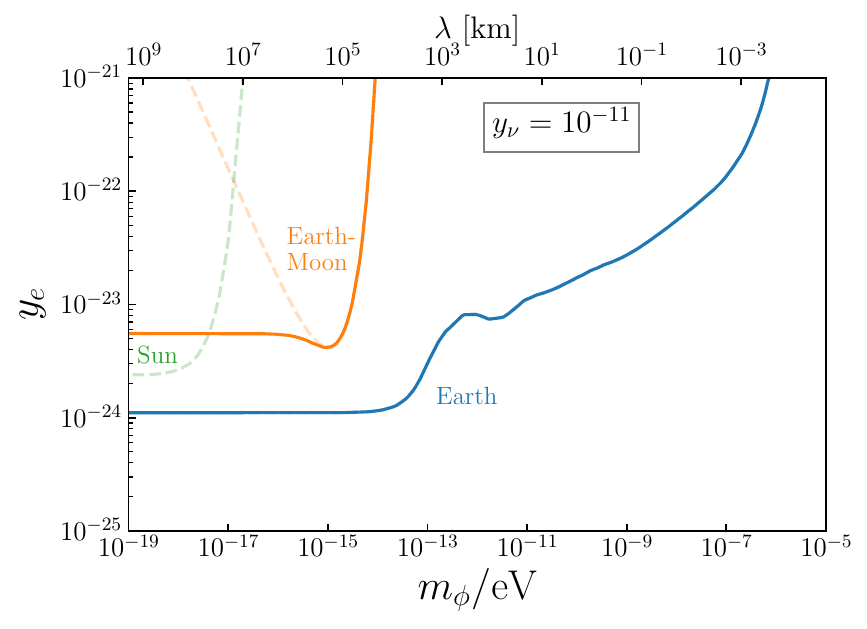}

\includegraphics[width=0.5\textwidth]{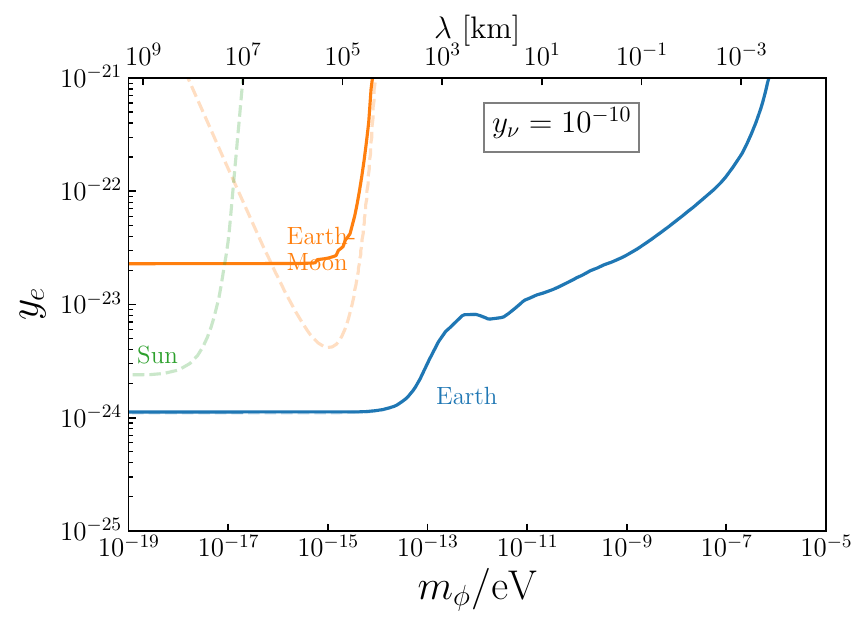}\includegraphics[width=0.5\textwidth]{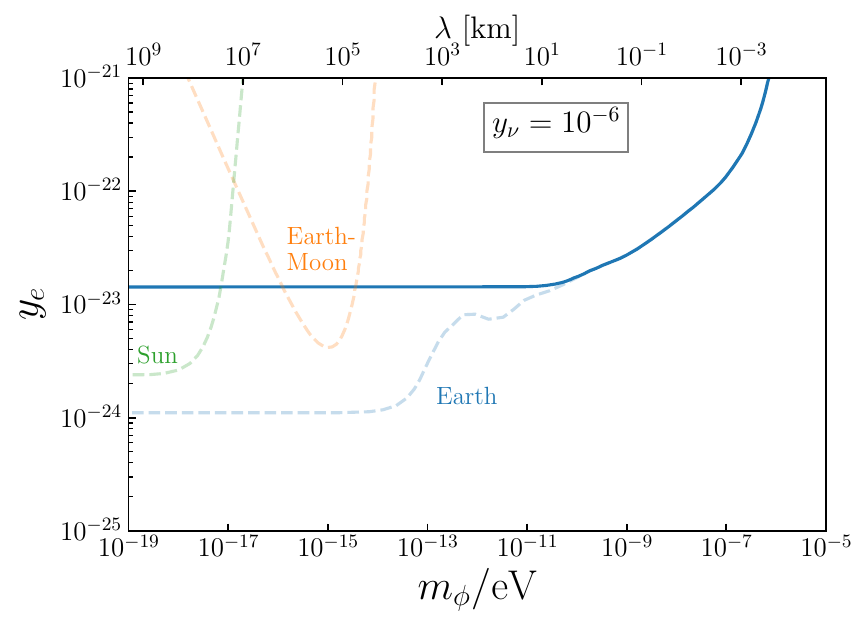}\caption{The impact of the C$\nu$B medium effect on long-range force searches.
The top left panel shows the experimental bounds on $y_{e}$ without
the C$\nu$B medium effect. The subsequent panels demonstrate the
variation of these bounds due to the C$\nu$B medium effect,  with
the dashed and solid lines representing the bounds before and after
including such an effect, respectively. The $x$-axis on the top of
each panel indicates the wavelength scale $\lambda\equiv1/m_{\phi}$.
\label{fig:result}}
\end{figure}

In the literature, these measurements have been used to use to set
constraints on the standard Yukawa potential in Eq.~\eqref{eq:-9}.
Typically for $m_{\phi}$ ranging from $10^{-14}$ eV (corresponding
to the length scale of the Earth radius) to $10^{-18}$ eV (corresponding
to the length scale of Earth-Sun distance), the experimental bounds
on $y_{\psi}$ for $\psi\in\{e^{-},\ p,\ n\}$ vary from $10^{-22}$
to $10^{-24}$---see the top left panel of Fig.~\ref{fig:result}. % e.g. Ref.~\cite{Heeck:2014zfa} or

Let us first estimate the medium effect caused by $\phi$ interacting
with electrons in the medium, which leads to a mass correction of
the order of
\begin{equation}
y_{e}\left(\frac{n_{e}}{m_{e}}\right)^{1/2}\sim10^{-22}\ \text{eV}\cdot\left(\frac{y_{e}}{10^{-24}}\right)\cdot\left(\frac{\rho}{5\ \text{g}/\text{cm}^{3}}\right)^{1/2},\label{eq:-34}
\end{equation}
where $\rho$ denotes the matter density and $5\ \text{g}/\text{cm}^{3}$
is the average density of the Earth. In Eq.~\eqref{eq:-34} we have
assumed $n_{e}=n_{p}\approx n_{n}$ so that $n_{e}\approx0.5\rho/m_{p}$.
Obviously this medium effect is negligibly small for the experiments
considered here. 

Next, we estimate the medium effect caused by the C$\nu$B which is
ubiquitous in the entire universe and according to the standard cosmological
model has the number density $56\ \nu/{\rm cm}^{3}+56\ \overline{\nu}/{\rm cm}^{3}$
for Dirac neutrinos or $112\ \nu/{\rm cm}^{3}$ for Majorana neutrinos.
The medium effect due to the C$\nu$B can be substantially enhanced by the smallness
of neutrino masses. Besides, the Yukawa coupling $y_{\nu}$ for very
light $\phi$ can be much greater than $y_{e}$ as known bounds on
$y_{\nu}$ are much less restrictive than those on $y_{e}$. Theoretically,
$y_{\nu}\gg y_{e}$ can be well motivated from models that introduce
light mediators via the neutrino portal~\cite{Xu:2020qek,Chauhan:2020mgv,Chauhan:2022iuh}. 

For nonrelativistic cosmic relic neutrinos ($m_{\nu}\gg1.6\times10^{-4}$
eV), the mass correction can be estimated as follows:
\begin{equation}
y_{\nu}\left(\frac{n_{\nu}}{m_{\nu}}\right)^{1/2}\sim10^{-16}\ \text{eV}\cdot\left(\frac{y_{\nu}}{10^{-10}}\right)\cdot\left(\frac{n_{\nu}}{56/\text{cm}^{3}}\right)^{1/2}\left(\frac{0.1\text{eV}}{m_{\nu}}\right)^{1/2}.\label{eq:-34-1}
\end{equation}

For relativistic neutrinos, according to Eq.~\eqref{eq:-13}, the
mass correction should be independent of $m_{\nu}$:
\begin{equation}
y_{\nu}\left(\frac{\tilde{n}_{\nu}}{m_{\nu}}\right)^{1/2}\sim10^{-15}\ \text{eV}\cdot\left(\frac{y_{\nu}}{10^{-10}}\right)\cdot\left(\frac{n_{\nu}}{56/\text{cm}^{3}}\right)^{1/3}.\label{eq:-34-2}
\end{equation}

The mass corrections of $10^{-15}$ or $10^{-16}$ eV in Eqs.~\eqref{eq:-34-1}
and \eqref{eq:-34-2}  are comparable to the inverse of the geometrical
sizes of the celestial bodies being considered.   Hence the above
estimates imply that the C$\nu$B medium effect can be considerably
large in the aforementioned experiments. 

Indeed, as is shown in Fig.~\ref{fig:result}, for a rather broad
range of $y_{\nu}$, the C$\nu$B may significantly alter some of
the experimental bounds on long-range forces. In this figure, we select
three representative experiments, the Moscow experiment using the
Sun as the attractor (green curves), the E\"ot-Wash experiment using
the Earth as the attractor (blue), and LLR which measures the varying
Earth-Moon distance (orange). The experimental bounds without including
medium effects are taken from Ref.~\cite{Wagner:2012ui}. We recast
the bounds into those presented in Fig.~\ref{fig:result} with significant
C$\nu$B medium effects included. Since the C$\nu$B is a homogeneous
distribution, the main effect is the mass correction in Eq.~\eqref{eq:-5-3}.
The C$\nu$B could also contribute to the source term on the right-hand
side of Eq.~\eqref{eq:-5-2}, but due to the homogeneity this contribution
only causes an constant shift of $\phi$, corresponding to adding a constant to the effective potential. Hence it can be omitted. 

More specifically, we recast the bounds as follows. For each given
$y_{\nu}$ and $m_{\phi}$, we can compute the effective mass $\tilde{m}_{\phi}$
according to Eq.~\eqref{eq:-5-3}, assuming that the mass corrections due to
$y_{\psi}$ with $\psi\in\{e^{-},\ p,\ n\}$ are negligible, which
has been justified by Eq.~\eqref{eq:-34}. Since the mass correction caused
by $y_{\nu}$ is independent of $r$, the new experimental bounds
on $y_{\psi}$ including the medium effect can be obtained by matching
$y_{\psi}^{(\text{new})}(m_{\phi})=y_{\psi}^{(\text{old})}(\tilde{m}_{\phi})$,
where $y_{\psi}^{(\text{new})}(m_{\phi})$ and $y_{\psi}^{(\text{old})}(m_{\phi})$
denote the new and old bounds with and without the medium effect included,
respectively. Note that this is valid only when the effective mass
is independent of $r$. If the C$\nu$B is not homogeneous at local
scales or if the mass corrections due to $y_{\psi}$ with $\psi\in\{e^{-},\ p,\ n\}$
becomes significant, we should numerically solve Eq.~\eqref{eq:-5-2} to obtain
the potential, with the geometry of the actual matter distribution
taken into account, and confront it with the experimental data to
obtain the actual bound.

In Fig.~\ref{fig:result}, we gradually increase the coupling $y_{\nu}$
from $0$ to $10^{-6}$ to show the impact of the C$\nu$B medium
effect on long-range force searches, assuming relativistic C$\nu$B.
For nonrelativistic C$\nu$B, the results are similar and the difference
can be absorbed by rescaling $y_{\nu}$ as follows:  
\begin{equation}
y_{\nu}\to18.5\times\left(\frac{m_{\nu}}{0.1\ \text{eV}}\right)^{1/2}y_{\nu}\thinspace.\label{eq:-35}
\end{equation}

In the upper right panel, we set $y_{\nu}=10^{-13}$ which according
to Eq.~\eqref{eq:-34-2} leads to a mass correction of $10^{-18}$
eV, corresponding to the inverse of the Earth-Sun distance. Consequently,
 the green curve in the top right panel is significantly altered and
becomes a weaker bound. It is interesting to note that the LLR bound
(orange curve) in this panel becomes more constraining for $m_{\phi}\lesssim10^{-18}$
eV. This is because the LLR bound is essentially a constraint on deviations
from the inverse square law. Without the medium effect, sufficiently
light mediators cannot be effectively  constrained by LLR since the
new force in the massless limit also follows the inverse square law.
With the medium effect, the effective mass is dominated by the medium-induced
mass if the vacuum mass $m_{\phi}$ is small. So LLR sets a bound
on $y_{e}$ independent of $m_{\phi}$ in this regime. 

Further increasing $y_{\nu}$, some bounds can be substantially alleviated
or even entirely eliminated. For example, for $y_{\nu}\gtrsim5\times10^{-13}$
and $3\times10^{-10}$, the green and orange bounds would disappear
respectively, as the medium effect renders the forces short-range
with respect to the relevant distances. One may ask to what extent
these long-range force bounds can be alleviated by continuing increasing
$y_{\nu}$. In fact, when $y_{\nu}$ is too large, the $\phi$ field
would thermalize via neutrino scattering in the early universe, modifying
the BBN observables and the cosmological effective number of relativistic neutrino species, $N_{{\rm eff}}$. According
to Refs.~\cite{Huang:2017egl,Li:2023puz},  for light scalars predominantly 
coupled to neutrinos, the cosmological upper bound on $y_{\nu}$ is
about $10^{-5}$. Therefore, we stop increasing $y_{\nu}$ at $10^{-6}$
in Fig.~\ref{fig:result}. In addition to the cosmological bound, there are other bounds on $y_{\nu}$ from constraints on neutrino self-interactions~\cite{Blinov:2019gcj,Deppisch:2020sqh,Brdar:2020nbj,Berryman:2022hds,Esteban:2021tub,Chang:2022aas,Wu:2023twu}. Typically laboratory bounds on $y_{\nu}$ can reach $10^{-3}$ to $10^{-5}$. 

From Fig.~\ref{fig:result}, we conclude that the C$\nu$B medium
effect on long-range force searches can fully eliminate the bounds
derived from LLR and the Sun, and alleviate the bound from the Earth
by about one order of magnitude.

\section{Conclusions  \label{sec:Conclusions}}

 Long-range force searches generally set the most constraining bounds
on ultra-light bosons that interact with normal matter. In this work,
we assume that such bosons are also coupled to neutrinos and investigate
whether the presence of cosmic neutrino background (C$\nu$B) could
have an impact on long-range force search experiments which are soaked
in this ubiquitous background. We find that  the medium effect of
C$\nu$B can significantly shorten the interaction ranges of long-range
forces, due to coherent forwarding scattering of the force mediator
with neutrinos. Consequently, the experimental bounds on long-range
forces can be altered by the C$\nu$B with sizable couplings of the
mediator to neutrinos.

Specifically, for experiments sensitive to new forces with interaction
 ranges longer than  $\sim10^{8}$ km (the Sun-Earth distance, corresponding
to a mediator mass of $\sim10^{-18}$ eV) or $\sim10^{5}$ km (the
Earth-Moon distance), the C$\nu$B could effectively screen the forces
from detection with the neutrino coupling greater than $5\times10^{-13}$
or $3\times10^{-10}$, respectively, causing such experimental bounds
to be entirely eliminated. For experiments utilizing the Earth as
an attractor,  the bounds can be substantially alleviated by about
one order of magnitude with cosmologically allowed values of the neutrino
coupling.    

Our results can also be used in the studies of neutrino oscillations
with long-range forces---see e.g.~\cite{Joshipura:2003jh,Grifols:2003gy,Gonzalez-Garcia:2006vic,Heeck:2010pg,Bustamante:2018mzu,Smirnov:2019cae,Babu:2019iml,Singh:2023nek}. Besides, future gravitational wave experiments such as LIGO, VIRGO,
Einstein Telescope, LISA, NANOGrav, etc., will be able to probe a
multitude of long-range force effects in both astrophysical and terrestrial
environments. For instance, neutron star mergers may provide a novel
avenue to long-range force searches, since the observed gravitational
waves could be altered by extra forces between two neutron stars as
well as extra radiations~\cite{Kopp:2018jom,Dror:2019uea}. The high density of a neutron
star could  modify such forces due to the medium effect discussed
in this work. On the other hand, it is also possible that the extra
radiations are enhanced or suppressed by the mass correction caused
by the surrounding cosmic neutrino background. We leave dedicated
studies on these possibilities for future work.
% We leave a dedicated study on this subject for future work. 

\begin{acknowledgments}
We thank Xuheng Luo for early discussions that inspired this work.
The work of GC is supported by the U.S. Department of Energy under the award number DE-SC0020250 and DE-SC0020262. The work of XJX is supported in part
by the National Natural Science Foundation of China under grant No.~12141501
and also by the CAS Project for Young Scientists in Basic Research
(YSBR-099). 
\end{acknowledgments}

\appendix

\section{Some minus signs\label{sec:minus-signs}}

In this appendix, we shall  clarify a few minus signs important to
our analysis. 

\subsection{Minus signs responsible for repulsive and attractive forces}

If the scalar mediator $\phi$ is replaced by a vector mediator $\phi^{\mu}$,
the long-range force between two $\psi$ particles becomes repulsive
instead of attractive. This change arises essentially from the differences
in their kinetic and mass terms. The mass term of a vector boson is
positive, implying a sign flip in the mass term:
\begin{equation}
-\frac{1}{2}m_{\phi}^{2}\phi^{2}\ \to\ +\frac{1}{2}m_{\phi}^{2}\phi^{\mu}\phi_{\mu}\thinspace.\label{eq:-15}
\end{equation}
On the other hand, the kinetic terms also differ by a minus sign:
\begin{equation}
\frac{1}{2}\left(\partial\phi\right)^{2}=-\frac{1}{2}\phi\partial^{2}\phi\ \to\ -\frac{1}{4}F^{\mu\nu}F_{\mu\nu}=\frac{1}{2}\phi^{\mu}\partial^{2}\phi_{\mu}\thinspace,\label{eq:-16}
\end{equation}
where $F^{\mu\nu}\equiv\partial^{\mu}\phi^{\nu}-\partial^{\nu}\phi^{\mu}$.
Consequently, the EOM in Eq.~\eqref{eq:-3} changes to 
\begin{equation}
-(\partial^{2}+m_{\phi}^{2})\phi^{\mu}+y_{\psi}\overline{\psi}\gamma^{\mu}\psi=0\thinspace.\label{eq:-17}
\end{equation}
Due to the minus sign in Eq.~\eqref{eq:-17}, the force becomes repulsive
between two $\psi$ particles.  One may wonder if the sign of $y_{\psi}$
in the Lagrangian \eqref{eq:-1} is relevant. This sign is unimportant
because the effective potential and the mass correction due to the
medium effect are all proportional to $y_{\psi}^{2}$. So flipping
the sign of $y_{\psi}$ has no physical consequences.

\subsection{When do antiparticle and particle contributions cancel?}

Antiparticles may introduce some extra minus signs to the calculations.
In Eqs.~\eqref{eq:-5}, \eqref{eq:-6}  and \eqref{eq:-5-1}, only
particles (no antiparticles) are included. In the presence of antiparticles,
one may be concerned with possible cancellations between antiparticles
and particles.

Let us first inspect the role of antiparticles in the ensemble averages
$\langle\overline{\psi}\psi\rangle$ and $\langle\overline{\psi}\gamma^{\mu}\psi\rangle$,
which determine the scalar and vector field strengths of $\phi$ ($\phi^{\mu}$),
respectively. Under charge conjugation (the $C$ transformation),
we have 
\begin{align}
\langle\overline{\psi}\gamma^{\mu}\psi\rangle & \overset{C}{\to}-\langle\overline{\psi}\gamma^{\mu}\psi\rangle\thinspace,\label{eq:-18}\\
\langle\overline{\psi}\psi\rangle & \overset{C}{\to}\langle\overline{\psi}\psi\rangle\thinspace,\label{eq:-19}
\end{align}
which implies  that $\langle\overline{\psi}\gamma^{\mu}\psi\rangle$
and $\langle\overline{\psi}\psi\rangle$ should be antisymmetric and
symmetric under the particle-antiparticle interchange: 
\begin{align}
\langle\overline{\psi}\gamma^{\mu}\psi\rangle & =\int\frac{d^{3}\mathbf{p}}{(2\pi)^{3}}\frac{p^{\mu}}{E_{\mathbf{p}}}\left[f_{+}(\mathbf{p})-f_{-}(\mathbf{p})\right],\label{eq:-20}\\
\langle\overline{\psi}\psi\rangle & =\int\frac{d^{3}\mathbf{p}}{(2\pi)^{3}}\frac{m_{\psi}}{E_{\mathbf{p}}}\left[f_{+}(\mathbf{p})+f_{-}(\mathbf{p})\right],\label{eq:-21}
\end{align}
where $f_{+}$ and $f_{-}$ denote the phase space distribution functions
of $\psi$ particles and antiparticles, respectively. Note that here
$\frac{d^{3}\mathbf{p}}{(2\pi)^{3}}\frac{1}{E_{\mathbf{p}}}$ is Lorentz
invariant. 

Eqs.~\eqref{eq:-20} and \eqref{eq:-21} are  obtained simply from
the argument of the $C$ symmetry. They can also be obtained by explicitly
computing the ensemble averages (i.e.~performing Wick contractions
of creation and annihilation operators) using the following identities:
\begin{align}
\overline{u}\gamma^{\mu}u & =2p^{\mu}\thinspace,\ \ \overline{v}\gamma^{\mu}v=2p^{\mu}\thinspace,\label{eq:-22}\\
\overline{u}u & =2m_{\psi}\thinspace,\ \ \overline{v}v=-2m_{\psi}\thinspace,\label{eq:-23}
\end{align}
where $u$ and $v$ denote the particle and antiparticle solutions
of the Dirac equation in momentum space, i.e.~$(\slashed{p}-m_{\psi})u=0$
and $(\slashed{p}+m_{\psi})v=0$. Note that the $u\leftrightarrow v$
interchange causes a minus sign for the scalar case in Eq.~\eqref{eq:-23},
as opposed to Eq.~\eqref{eq:-19} or ~\eqref{eq:-21} where the scalar
product is symmetric under the particle-antiparticle interchange.
This minus sign is canceled by interchanging anti-commuting fermionic
creation and annihilation operators. For the vector case, the same
reason accounts for the anti-symmetry in Eq.~\eqref{eq:-20} when
deriving it from Eq.~\eqref{eq:-22}. 

Finally, let us inspect the role of antiparticles in the mass correction
\eqref{eq:-6}. The medium-induced mass correction for the scalar
case arises from the evaluation of the following ensemble average:
\begin{equation}
\Delta m_{\phi}^{2}\phi^{2}\propto y_{\psi}^{2}\langle\overline{\psi}\phi\psi\overline{\psi}\phi\psi\rangle\thinspace,\label{eq:-24}
\end{equation}
which is obviously a $C$-even quantity according to Eq.~\eqref{eq:-19}.
Hence antiparticles contribute positively to Eq.~\eqref{eq:-6}, which
after including such a contribution should be
\begin{equation}
m_{\phi}^{2}\to m_{\phi}^{2}+y_{\psi}^{2}\frac{\tilde{n}_{\psi}+\tilde{n}_{\overline{\psi}}}{m_{\psi}}\thinspace.\label{eq:-6-1}
\end{equation}
As for the vector case, although $\overline{\psi}\gamma^{\mu}\psi$
is $C$-odd, the medium-induced mass correction $\langle\overline{\psi}\gamma\cdot\phi\psi\overline{\psi}\gamma\cdot\phi\psi\rangle$
is $C$-even. Hence the antiparticle contribution is also positive,
which is a known conclusion for the plasmon mass in
$e^{+}$$e^{-}$ plasma. 

\section{Derivation of the medium effect from coherent forward scattering
\label{sec:Rederivation}}

As mentioned in Sec.~\ref{subsec:The-medium-effect}, the medium
effect given by Eqs.~\eqref{eq:-6} and \eqref{eq:-10}, known from
the finite-temperature field theory, is also valid for non-thermal
distributions, because it fundamentally arises from coherent forward
scattering of $\phi$ with medium particles. In this appendix, we
rederive this medium effect using the theory of coherent forward scattering---similar
calculations for photon-dark photon or photon-axion conversions can
be found in Refs.~\cite{Li:2023vpv,Wu:2024fsf}. 

\begin{figure}
\centering

\includegraphics[width=0.49\textwidth]{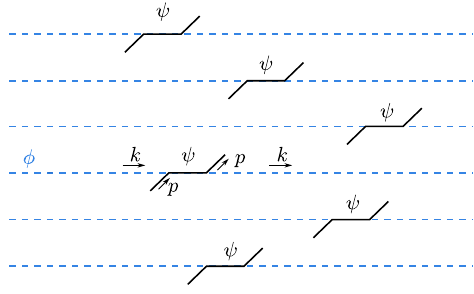}\caption{A schematic picture illustrating coherent forward scattering of $\phi$
with medium particles. The momenta of both $\phi$ and $\psi$ remain
unchanged after the scattering. As a consequence, all scattering amplitudes
can be added coherently. The combined medium effect slows down the
propagation of the $\phi$ field.  \label{fig:scat}}
\end{figure}

In coherent forward scattering, the medium particles do not change
their states after such scattering, implying that the amplitude of
scattering with each medium particle can be added coherently, as is
illustrated by Fig.~\ref{fig:scat}. The combined effect corresponds
to the mass correction in Eq.~\eqref{eq:-6}. 

Let us first compute the scattering of $\phi$ with a single $\psi$
particle, for which the amplitude reads
\begin{equation}
i{\cal M}=y_{\psi}^{2}\overline{u}(p)\left[\frac{i}{(p+k)\cdot\gamma-m_{\psi}}+\frac{i}{(p-k)\cdot\gamma-m_{\psi}}\right]u(p)\thinspace,\label{eq:-36}
\end{equation}
where $u$ and $\overline{u}$ denote the initial and final fermion
states; $p$ and $k$ denote the momenta of the fermion and the scalar,
respectively. The second term in square brackets accounts for the
interchange of the two Yukawa vertices. Applying on-shell conditions
for the fermion, $p^{2}=m_{\psi}^{2}$ and $\slashed{p}u=m_{\psi}u$,
we obtain
\begin{equation}
i{\cal M}=iy_{\psi}^{2}\overline{u}\left[\frac{4k^{2}m_{\psi}-4(p\cdot k)(k\cdot\gamma)}{k^{4}-4(k\cdot p)^{2}}\right]u\thinspace.\label{eq:-37}
\end{equation}

Next, we consider such scattering in a medium containing a large number
of $\psi$ particles, which form a background state denoted by $|\text{bkg}\rangle$.
In this case, we need to evaluate the following matrix element:

\begin{equation}
\langle {\rm bkg}|{\overline{\psi}}\phi {\psi}{\overline{\psi}}\phi {\psi} |{\rm bkg}  \rangle
=
\bcontraction{\langle} {\rm bkg} {|} {\overline{\psi}} {}
\bcontraction{\langle {\rm bkg}|{\overline{\psi}}\phi}  {\psi}   {} {\overline{\psi}} {}
\bcontraction{\langle {\rm bkg}|{\overline{\psi}}\phi {\psi}{\overline{\psi}}\phi}  {\psi}   {|} {\rm bkg}   
\langle {\rm bkg}|{\overline{\psi}}\phi {\psi}{\overline{\psi}}\phi {\psi} |{\rm bkg}  
\rangle
+\text{vertex interchange}\,,
\label{eq:contract}
\end{equation}where the first and third Wick contractions correspond to $\overline{u}$
and $u$ in Eq.~\eqref{eq:-36}, and the second contraction corresponds
to the fermion propagator in Eq.~\eqref{eq:-36}, respectively. The
``vertex interchange'' term corresponds to the second term in Eq.~\eqref{eq:-36}.
Therefore, Eq.~\eqref{eq:-37} implies
\begin{align}
\langle{\rm bkg}|\overline{\psi}\phi\psi\overline{\psi}\phi\psi|{\rm bkg}\rangle & =\phi^{2}\langle\text{bkg}|\overline{\psi}\thinspace\frac{4k^{2}m_{\psi}-4(p\cdot k)(k\cdot\gamma)}{k^{4}-4(k\cdot p)^{2}}\thinspace\psi|\text{bkg}\rangle\nonumber \\
 & =\phi^{2}\int\frac{d^{3}\mathbf{p}}{(2\pi)^{3}}\frac{f_{+}(\mathbf{p})}{E_{\mathbf{p}}}\frac{4k^{2}m_{\psi}^{2}-4(p\cdot k)(k_{\mu}p^{\mu})}{k^{4}-4(k\cdot p)^{2}}\thinspace,\label{eq:-38}
\end{align}
where in the second step we have used Eqs.~\eqref{eq:-20} and \eqref{eq:-21}.
For simplicity, we assume that the background does not contain antiparticles.
The antiparticle contribution can be readily included by performing
charge conjugation on the final result. 

For applications in generating effective potentials, one can take
the limit $k^{2}\to0$. Then Eq.~\eqref{eq:-38} implies 
\begin{equation}
y_{\psi}^{2}\langle{\rm bkg}|\overline{\psi}\phi\psi\overline{\psi}\phi\psi|{\rm bkg}\rangle=y_{\psi}^{2}\phi^{2}\int\frac{d^{3}\mathbf{p}}{(2\pi)^{3}}\frac{f_{+}(\mathbf{p})}{E_{\mathbf{p}}}=y_{\psi}^{2}\phi^{2}\frac{\tilde{n}_{\psi}}{m_{\psi}}\thinspace,\label{eq:-39}
\end{equation}
which is exactly the mass correction  in Eq.~\eqref{eq:-6}.

\bibliographystyle{JHEP}
\bibliography{ref}

\end{document}